%% file: main.tex
\documentclass[runningheads]{llncs}
\usepackage[T1]{fontenc}
\usepackage{graphicx}
\usepackage{algorithm}
\usepackage[noend]{algpseudocode}
\usepackage{algorithmicx}
\usepackage{graphicx}
\usepackage{textcomp}
\usepackage{xcolor}
\usepackage{calc}        
\usepackage{amsmath,amsfonts}
\usepackage{marvosym}
\usepackage{venndiagram}
\usepackage{listings}
\usepackage{enumitem}
\usepackage{tikz}
\usetikzlibrary{shapes,backgrounds}
\usepackage{framed}
\usepackage{float}
\usepackage{multirow}
\usepackage{array}
\usepackage{cite}
\usepackage{xspace}
\usepackage[breakable]{tcolorbox}
\usepackage{subfigure}
\usepackage[colorlinks=true,linkcolor=blue, citecolor=blue]{hyperref}
\usepackage{cleveref}
\usepackage{comment}
\usepackage{url}
\usepackage{booktabs}

\algrenewcommand\algorithmicrequire{\textbf{Input:}}
\algrenewcommand\algorithmicensure{\textbf{Output:}}

\AtBeginDocument{%
  }

\usepackage{xcolor}
\usepackage{listings}
\newcommand\code[1]{{\tt\small #1}}
\definecolor{dkgreen}{rgb}{0,0.3,0}
\definecolor{cmtgreen}{rgb}{0,0.6,0}
\definecolor{ltblue}{rgb}{0,0.4,0.4}
\definecolor{dkviolet}{rgb}{0.3,0,0.5}
\definecolor{dkblue}{rgb}{0,0.2,0.2}
\lstdefinelanguage{Coq}{ 
    mathescape=true,
    texcl=false, 
    escapeinside={(@}{@)},
    morekeywords=[1]{Section, Module, End, Require, Import, Export,
        Variable, Variables, Parameter, Parameters, Axiom, Hypothesis,
        Hypotheses, Notation, Local, Tactic, Reserved, Scope, Open, Close,
        Bind, Delimit, Definition, Let, Ltac, Fixpoint, CoFixpoint, Add,
        Morphism, Relation, Implicit, Arguments, Unset, Contextual,
        Strict, Prenex, Implicits, Inductive, CoInductive, Record,
        Structure, Canonical, Coercion, Context, Class, Global, Instance,
        Program, Infix, Theorem, Lemma, Corollary, Proposition, Fact,
        Remark, Example, Proof, Goal, Save, Qed, Defined, Hint, Resolve,
        Rewrite, View, Search, Show, Print, Printing, All, Eval, Check,
        Projections, inside, outside, Def},
    morekeywords=[2]{forall, exists, exists2, fun, fix, cofix, struct,
        match, with, end, as, in, return, let, if, is, then, else, for, of,
        nosimpl, when},
    morekeywords=[3]{Type, Prop, Set, true, false, option},
    morekeywords=[4]{pose, set, move, case, elim, apply, clear, hnf,
        intro, intros, generalize, rename, pattern, after, destruct,
        induction, using, refine, inversion, injection, rewrite, congr,
        unlock, compute, ring, field, fourier, replace, fold, unfold,
        change, cutrewrite, simpl, have, suff, wlog, suffices, without,
        loss, nat_norm, assert, cut, trivial, revert, bool_congr, nat_congr,
        symmetry, transitivity, auto, split, left, right, autorewrite},
    morekeywords=[5]{by, done, exact, reflexivity, tauto, romega, omega,
        assumption, solve, contradiction, discriminate},
    morekeywords=[6]{do, last, first, try, idtac, repeat},
    morecomment=[s]{(*}{*)},
    showstringspaces=false,
    morestring=[b]",
    morestring=[d]’,
    tabsize=3,
    extendedchars=false,
    sensitive=true,
    breaklines=false,
    basicstyle=\fontsize{9.5pt}{11.4pt}\selectfont\ttfamily,
    captionpos=b,
    columns=[l]flexible,
    identifierstyle={\ttfamily\color{black}},
    keywordstyle=[1]{\bfseries\ttfamily\color{dkviolet}},
    keywordstyle=[2]{\bfseries\ttfamily\color{dkgreen}},
    keywordstyle=[3]{\bfseries\ttfamily\color{ltblue}},
    keywordstyle=[4]{\bfseries\ttfamily\color{dkblue}},
    keywordstyle=[5]{\bfseries\ttfamily\color{dkred}},
    stringstyle=\ttfamily,
    commentstyle={\bfseries\ttfamily\color{cmtgreen}},
    literate=
    {\\forall}{{\color{dkgreen}{$\forall\;$}}}1
    {\\exists}{{$\exists\;$}}1
    {<-}{{$\leftarrow\;$}}1
    {=>}{{$\Rightarrow\;$}}1
    {==}{{\code{==}\;}}1
    {==>}{{\code{==>}\;}}1
    {->}{{$\rightarrow\;$}}1
    {<->}{{$\leftrightarrow\;$}}1
    {<==}{{$\leq\;$}}1
    {\#}{{$^\star$}}1 
    {\\o}{{$\circ\;$}}1 
    {\@}{{$\cdot$}}1 
    {\/\\}{{$\wedge\;$}}1
    {\\\/}{{$\vee\;$}}1
    {++}{{\code{++}}}1
    {~}{{$\sim$}}1
    {\@\@}{{$@$}}1
    {\\mapsto}{{$\mapsto\;$}}1
    {\\hline}{{\rule{\linewidth}{0.5pt}}}1
}[keywords,comments,strings]

\lstdefinelanguage{PlainText}{ 
    mathescape=true,
    texcl=false, 
    escapeinside={(@}{@)},
    morecomment=[s]{(*}{*)},
    showstringspaces=false,
    morestring=[b]",
    morestring=[d]’,
    tabsize=3,
    extendedchars=false,
    sensitive=true,
    breaklines=false,
    basicstyle=\fontsize{5.5pt}{7.5pt}\selectfont\ttfamily,
    captionpos=b,
    columns=[l]flexible,
    identifierstyle={\ttfamily\color{black}},
    keywordstyle=[1]{\bfseries\ttfamily\color{dkviolet}},
    keywordstyle=[2]{\bfseries\ttfamily\color{dkgreen}},
    keywordstyle=[3]{\bfseries\ttfamily\color{ltblue}},
    keywordstyle=[4]{\bfseries\ttfamily\color{dkblue}},
    keywordstyle=[5]{\bfseries\ttfamily\color{dkred}},
    stringstyle=\ttfamily,
    commentstyle={\bfseries\ttfamily\color{dkgreen}},
    literate=
    {\\forall}{{\color{dkgreen}{$\forall\;$}}}1
    {\\exists}{{$\exists\;$}}1
    {<-}{{$\leftarrow\;$}}1
    {=>}{{$\Rightarrow\;$}}1
    {==}{{\code{==}\;}}1
    {==>}{{\code{==>}\;}}1
    {->}{{$\rightarrow\;$}}1
    {<->}{{$\leftrightarrow\;$}}1
    {<==}{{$\leq\;$}}1
    {\#}{{$^\star$}}1 
    {\\o}{{$\circ\;$}}1 
    {\@}{{$\cdot$}}1 
    {\/\\}{{$\wedge\;$}}1
    {\\\/}{{$\vee\;$}}1
    {++}{{\code{++}}}1
    {~}{{$\sim$}}1
    {\@\@}{{$@$}}1
    {\\mapsto}{{$\mapsto\;$}}1
    {\\hline}{{\rule{\linewidth}{0.5pt}}}1
}[keywords,comments,strings]

\algrenewcommand\algorithmicrequire{\textbf{Input:}}
\algrenewcommand\algorithmicensure{\textbf{Output:}}

\newcommand{\approach}{\textsc{Strat2Rocq}}
\newcommand{\coqart}{\textsc{Coq-Art}}
\newcommand{\extlib}{\textsc{Ext-Lib}}
\newcommand{\vfa}{\textsc{Vfa}}
\newcommand{\compcert}{\textsc{CompCert}}
\newcommand{\theoremsum}{2,394}
\newcommand{\evalres}{13.41\%}
\newcommand{\evalllm}{4.00\%}

\renewcommand{\paragraph}[1]{\smallskip\noindent\textbf{#1.}}

\newtcolorbox{rqbox}{breakable,left=4pt,right=4pt,top=4pt,bottom=4pt}
\begin{document}
\title{Proof Strategy Extraction from LLMs for\\ Enhancing Symbolic Provers}

%
%
\author{Jian Fang \and
Yican Sun \and
Yingfei Xiong\textsuperscript{(\Letter)}}
\authorrunning{J. Fang et al.}

\institute{Key Laboratory of High Confidence Software Technologies (Peking University), Ministry of Education; School of Computer Science, Peking University, Beijing, China
\email{fangjian@stu.pku.edu.cn} \\
\email{\{sycpku,xiongyf\}@pku.edu.cn}}
\maketitle              
\begin{abstract}
One important approach to software verification is interactive theorem proving. However, writing formal proofs often requires substantial human effort, making proof automation highly important. Traditionally, proof automation has relied on symbolic provers. Recently, large language models (LLMs) have demonstrated strong capabilities in theorem proving, complementing symbolic provers. Nonetheless, prompting LLMs can be expensive and may pose security risks for confidential codebases. As a result, purely symbolic approaches remain important even in the LLM era, as they are cost-effective, secure, and complement the strengths of LLMs.

Motivated by these considerations, we pose a new research question: can the internal proof strategies of LLMs be extracted to enhance the capabilities of symbolic provers? As an initial step, we introduce \approach{}. In an offline stage, \approach{} extracts proof strategies from LLMs and formalizes them as lemmas in Rocq. In an online stage, given a theorem to be proved, \approach{} augments the proof context with these extracted lemmas, enabling CoqHammer to leverage the LLM-derived strategies for more effective automated proving.
Our evaluation demonstrates that, on open-source Rocq projects for software verification, \approach{} enhances the success rate of CoqHammer by \evalres{}. 
A side discovery is that the extracted lemmas are also beneficial to LLM proof agents, improving the success rate of an LLM proof agent by \evalllm{}.

\keywords{Software Verification \and Large Language Models \and Proof Strategy Extraction \and Symbolic Provers.}
\end{abstract}

\input{intro}
\input{overview}
\input{evaluation}
\input{related}
\input{discussion}

\bibliographystyle{splncs04}
\bibliography{mybibliography}

\end{document}

%% file: intro.tex
\section{Introduction}
\label{sec:introduction}

Software verification is a critical area in software engineering, aiming to ensure that a software system never performs unsafe operations or exhibits undesirable behavior~\cite{DBLP:conf/sosp/KleinEHACDEEKNSTW09,DBLP:conf/itp/KrebbersLW14}.
One important approach to software verification is interactive theorem proving (ITP), which has been successfully applied to build the certified C compiler~\cite{DBLP:conf/itp/KrebbersLW14}, micro operating system kernel~\cite{DBLP:conf/sosp/KleinEHACDEEKNSTW09}, and cryptographic systems~\cite{protzenko2020evercrypt}. 
However, writing formal proofs often requires a substantial amount of effort. 
In seL4~\cite{DBLP:conf/sosp/KleinEHACDEEKNSTW09}, a verified OS kernel, the effort in writing the proof code is approximately 10 times as large as that of implementing the kernel. 



Recently, large language models (LLMs) have demonstrated significant capabilities in proof automation, effectively complementing traditional symbolic approaches~\cite{11029818,DBLP:journals/corr/abs-2410-19940,DBLP:conf/kbse/LuD024}. As shown in recent studies, proof agents, which integrate LLMs with ITPs, can achieve state-of-the-art performance.

However, these hybrid proof agents cannot fully replace purely symbolic approaches. 
LLMs typically require substantial computational resources, making them unsuitable for scenarios with limited hardware availability~\cite{j2024finetuningllmenterprise,han2025tokenbudgetawarellmreasoning}.
This drawback is especially relevant for many safety-critical software vendors, which are often small companies with constrained budgets. 
Furthermore, their codebases are often confidential, preventing the use of external LLM services.
In contrast, symbolic proof automation systems can be executed efficiently on a personal computer, without requiring significant computational infrastructure or posing risks of information leakage.


\paragraph{Research Question}
Given that symbolic provers remain important even in the era of LLMs--and indeed complement LLMs--we pose the following research question:
\begin{quote}
\begin{small}
\textit{Can we extract the proving capabilities of LLMs to enhance symbolic provers?}
\end{small}
\end{quote}

As noted in previous literature from cognitive science, human problem solving is often characterized by the combination of internally acquired strategies~\cite{NewellSimon1972,EricssonKintsch1995}. 
In the context of theorem proving, these proof strategies correspond to theorems and lemmas at which the human prover excels. Given that LLMs are trained on large corpora of human-created data, we hypothesize that LLMs adopt a similar paradigm when generating proofs.
For example, given the expression $a + b + c$, an LLM may be able to immediately infer that this expression is equal to $c + b + a$—or any other permutation—rather than applying a step-by-step sequence of transformations based on the laws of associativity and commutativity. 
If we can extract these proof strategies into a form accessible to symbolic provers, we may be able to improve the performance of symbolic provers.

The key question, then, is how to extract these proof strategies. 
Our approach is to inspect the proof trajectories generated by LLMs. If an LLM infers a conclusion directly from a set of premises without intermediate steps, we identify that inference as applying a proof strategy. 
We use lemmas to represent these strategies, since lemmas in modern formal proof languages are both expressive and accessible to provers. Hence, we design two separate stages for extraction and utilization of these strategies, namely \textit{offline lemma mining} and \textit{online proving}.

{\textbf{Stage 1: Offline lemma mining}}. In this stage, we are given a training set (a corpus of theorems). We scan each theorem in the training set and follow a prompt-and-formalize workflow as follows: i) First, we prompt the LLMs to prove each theorem. ii) Then, we extract strategies from the resulting proof trajectories and formalize these strategies as lemmas in a proof language (e.g., Rocq). This stage is executed only once.

{\textbf{Stage 2: Online theorem proving}}. After all lemmas have been extracted, we insert them into the context of new theorems to be proved, thereby making CoqHammer accessible to the internal knowledge distilled from the LLM. Unlike previous proof agents~\cite{11029818,DBLP:journals/corr/abs-2410-19940} where LLMs are inevitable in online proving, our approach only relies on a purely symbolic prover to prove a user-provided theorem.


As an initial validation of our idea, this paper introduces \approach{}, which extracts the proof strategies of LLMs as formalized lemmas in Rocq.
We select CoqHammer~\cite{DBLP:journals/jar/CzajkaK18}, the state-of-the-art symbolic prover in Rocq, as the target system to be improved.

\paragraph{Evaluation}
To evaluate \approach{}, we selected open-source Rocq projects focused on software verification, including:
\begin{itemize}
    \item \coqart{}~\cite{bertot2013interactive}: a textbook on Rocq in building simple certified programming languages and data structures.
    \item \extlib{}~\cite{coqextlib}: a community-driven Rocq library.
    \item \vfa{}~\cite{Appel:SF3}, the third volume of Software Foundations focusing on verified functional algorithms.
    \item \compcert{}~\cite{DBLP:conf/itp/KrebbersLW14}, the certified C compiler written in Rocq.
\end{itemize}
We collected all theorems from~\coqart{}, \extlib{} and \vfa{}, and sample 1,000 theorems from \compcert{}. As a result, we obtain a benchmark of \theoremsum{} formalized Rocq theorems. We applied 3-fold cross-validation and measured the success rate of CoqHammer both before and after granting it access to the new lemmas extracted by \approach{}. Our results demonstrate that \approach{} increases the proof success rate by \evalres{} in average. Furthermore, though our primary goal is to improve symbolic provers, the extracted lemmas are also beneficial to an LLM agent, improving the performance of an LLM agent by \evalllm{}.





\paragraph{Contribution}
In summary, the contributions of this paper are as follows:
\begin{itemize}
\item We propose a novel research problem: offline extracting proof strategies from large language models (LLMs) to enhance the performance of symbolic provers.
\item We propose an approach for extracting reusable lemmas by analyzing natural language proofs generated by LLMs, enabling the proof strategies of LLMs to be integrated into symbolic provers.
\item We implemented our method as \approach{} and empirically validated its effectiveness.
\end{itemize}

%% file: overview.tex
\section{Our Framework: \approach{}}
This section presents \approach{}, our framework for extracting knowledges from LLMs. 
Throughout this section, we will use the following running example for demonstration.

\label{sec:overview}
\begin{figure} 
\begin{lstlisting}[language=Coq, basicstyle=\scriptsize\ttfamily,xleftmargin=0em]
(* Monoid over the set A *)
Class Monoid (dot: A->A->A) (one: A) := {
  dot_assoc : forall x y z: A, dot x (dot y z) = dot (dot x y) z;
  one_left (@$\hspace{0.5em}$@): forall x: A, dot one x = x;
  one_right : forall x: A, dot x one = x  
}.
(* Generalized power operator the monoid M *)
Fixpoint power {M : Monoid A dot one} (a: A) (n: nat) :=
  match n with 
    | 0 => one
    | S p => dot a (power a p)
  end.
 (* The theorem to be proved, where dot and power are monoid operators *)
Theorem power_commute_with_x: forall (x: A) (n: nat), dot x (power x n) = dot (power x n) x.
\end{lstlisting}
\caption{Our Running Example}
\label{fig:running-ex}
\end{figure}

\begin{example}[Running Example]
\label{ex:running}
Consider the theorem \texttt{power\_commute\_with\_x} presented in \Cref{fig:running-ex}, which is extracted from the widely-used \coqart{} project. 
This theorem concerns a fundamental property of Monoids~\cite{10.1145/3122955.3122963,chiusano2014functional}. A Monoid is an abstract interface that captures the essence of many common data structures, such as natural numbers and lists, and finds applications in diverse programming scenarios including data aggregation, log system construction, and parser composition~\cite{chiusano2014functional,cole1995parallel,hutton1992higher}.
More precisely, as illustrated in \Cref{fig:running-ex}, a Monoid is defined over a base set \texttt{A} and is equipped with a multiplication operation \texttt{dot} that is associative and has a unit element \texttt{one}. Additionally, the power operator for a Monoid (the function \texttt{power} in \Cref{fig:running-ex}) is defined as a generalization of the familiar power operation on natural numbers.

With these definitions in place, our target theorem states the following: for every element $\mathtt{x} \in \mathtt{A}$, multiplying \texttt{x} to either the left or the right of \texttt{power x n} yields the same result; in both cases, the result is precisely \texttt{power x (n + 1)}.
Notably, CoqHammer is unable to prove this theorem. 
\qed
\end{example}

\begin{figure}[htbp]
\centering
\includegraphics[width=0.8\linewidth]{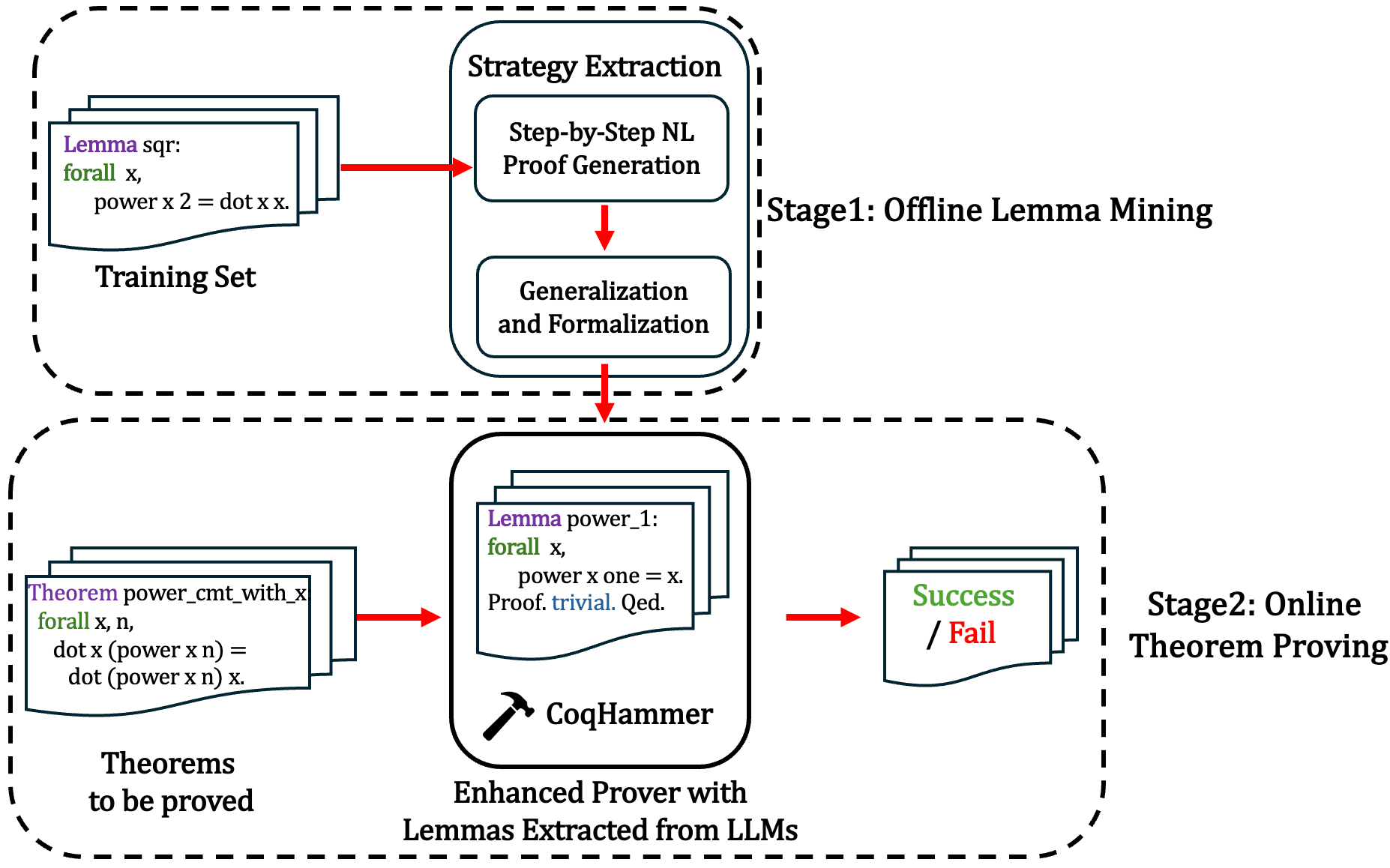}
\caption{The Workflow of \approach{}}
\label{fig:workflow}
\end{figure}

\paragraph{Workflow} The overall workflow is illustrated in \Cref{fig:workflow}. The proposed framework comprises two stages:
\begin{itemize}
\item \textbf{Stage 1: Offline Lemma Mining.}
In this stage, we extract internal proof strategies from LLMs over a given training set. For each theorem in the training set, we (1) generate a natural-language proof and subsequently (2) generalize and formalize each atomic proof step into a new lemma. These two steps are detailed in \Cref{sec:gen-nl-proof} and \Cref{sec:formalize}, respectively. This stage is performed only once.
\item \textbf{Stage 2: Online Theorem Proving.}  
After obtaining the formalized lemmas, they are inserted into the context of the user-provided new theorem to be proved. We describe this stage in \Cref{sec:proving}.
\end{itemize}

\begin{figure}
\begin{lstlisting}[frame=single,breaklines=true,breakindent=0pt,basicstyle=\small\ttfamily,escapechar=@,basicstyle=\scriptsize\ttfamily,xleftmargin=0em]
You are an expert at Rocq theorem proving. Please give a step-by-step natural language proof for the given theorem. 
Below, we further provide some helper information.
[Type Definitions] ...
[Function Definitions] ...
[Script So Far] ...
Finally, The given theorem is: ...
\end{lstlisting}
\caption{Detailed Prompts for Natural Language Proof Generation}
\label{fig:prompt-design-nl-proof-gen}
\end{figure}

\subsection{Step 1 of Stage 1: Natural-Language Proof Generation}
\label{sec:gen-nl-proof}


For each theorem \texttt{thm} in the training set, we begin by generating a natural-language (NL) proof. We adopt NL proofs rather than formal proofs because LLMs typically demonstrate stronger performance in producing NL proof. Moreover, when generating formal proofs, LLMs often first construct an NL proof and then formalize it. Consequently, the NL proof is likely to reflect the internal reasoning process of the LLM.

\begin{example}
\label{ex:train-set}
Continuing with \Cref{ex:running}, suppose the training set includes the following theorem, \texttt{sqr}, located in the file \texttt{Monoid.v}. The theorem states that for every element \texttt{x} in the monoid, its square is simply \texttt{dot x x}:
\begin{lstlisting}[language=Coq, basicstyle=\small\ttfamily]
Theorem sqr: forall x: A, power x 2 = dot x x.
\end{lstlisting}
In subsequent examples~(\Cref{ex:nl-proof,ex:formalize}), we will demonstrate that, after adding the lemmas formalized from the LLM-generated NL proof of the theorem \texttt{sqr}, CoqHammer becomes capable of proving the theorem in \Cref{ex:running}.
\qed
\end{example} 

Our prompt for generating the NL proof follows the template in \Cref{fig:prompt-design-nl-proof-gen}.
The prompt starts with a role description: ``You are an expert at Rocq theorem proving,'' followed by an articulation of the task, asking the LLM to generate the step-by-step NL proof for the given theorem. Our prompt also supplies essential background information, including:
\begin{itemize}
\item \textit{Type Definitions:} Definitions of all types referenced in \texttt{thm}.
\item \textit{Function Definitions:} Definitions of all functions used in \texttt{thm}. Supplying relevant types and functions helps the LLM better understand the theorem and generate higher-quality proofs.
\item \textit{Script So Far:} The content of the file containing \texttt{thm}, preceding the statement of \texttt{thm}. This information enables the LLM to accurately interpret the broader context in which the theorem is stated.
\end{itemize}
In response to our prompt, the LLM produces a step-by-step natural language proof for the current theorem.

%

\begin{example}
\label{ex:nl-proof}
Continuing with \Cref{ex:train-set}, as illustrated in \Cref{fig:prompt-design-nl-proof-gen}, the system prompt consists of the definitions of \texttt{Monoid} and \texttt{Power}, as well as all Rocq scripts in \texttt{Monoid.v} preceding the statement of the theorem \texttt{sqr}. The user prompt is simply the theorem statement of \texttt{sqr}. The response of LLM is presented below. 
\begin{lstlisting}[language=PlainText, basicstyle=\scriptsize\ttfamily,xleftmargin=3em,frame=single,xrightmargin=3em]
1) Introduce the only variable.
2) Simplify the definition of "power" at n = 2.
After this simplification, power x 2 = dot x (power x 1).
3) (@\textcolor{red}{\texttt{Simplify "power x 1":}} @) (@\textcolor{red}{\texttt{Now}}@) (@\textcolor{red}{\texttt{power x 1 = dot x (power x 0) = dot x one}.} @)
... Further steps omitted
\end{lstlisting}
Notably, there is a sub-step showing that \texttt{power x 1 = dot x one}. After formalization, this step becomes crucial for proving our target theorem~\Cref{ex:running}.
\end{example}

\subsection{Step 2 of Stage 1: Generalization and Formalization of the Natural-Language Proof Steps}
\label{sec:formalize}

\begin{figure}
\begin{lstlisting}[frame=single,breaklines=true,xleftmargin=0em,breakindent=0pt,basicstyle=\small\ttfamily,escapechar=@,linewidth=\linewidth,basicstyle=\scriptsize\ttfamily]
Based on the NL proof above, subsequently do the following tasks:
[Task 1] Generalize the proof steps above as general lemmas.
[Task 2] Formalize each lemma in Rocq, and provide the formal proof.
[Task 3] Return a Rocq script that includes all formalized lemmas and proofs.
\end{lstlisting}
\caption{Detailed Prompts for Generalization and Formalization}
\label{fig:prompt-design-generalize-and-formalize}
\end{figure}

To ensure that the extracted strategy is not restricted to a specific theorem in the training set, we prompt the LLM to generalize the properties identified in the natural-language proof. After obtaining these generalized strategies, we formalize them as lemmas in Rocq.

The latter task in this step is known also as \emph{autoformalization}~\cite{autoformalization}, which aims to automatically translate NL proofs into formal proofs. Autoformalization is widely regarded as an open challenge for LLMs~\cite{autoformalization,autoformalization-survey}. However, in our setting, we only require the formalization of each individual proof step, which substantially simplifies the autoformalization problem.

We reuse the same chat session described in \Cref{sec:gen-nl-proof} and append a new query using the prompt shown in \Cref{fig:prompt-design-generalize-and-formalize}. This prompt instructs the LLM, in a step-by-step manner, to generalize, formalize, and summarize the proof. An alternative design would be to first parse the NL proof into individual steps and then ask the LLM to generalize, formalize, and prove each step independently. We do not adopt this design for two reasons: (1) the LLM’s output may not adhere to a fixed format even when one is specified in the prompt, and parsing such NL output can lead to frequent failures; and (2) the steps in the NL proof often depend on one another, so even if parsing were successful, it would still be difficult to independently formalize each step.

In practice, LLMs often fail to produce correct lemmas and formal proofs in a single attempt. To mitigate this, we employ a standard proof agent that iteratively invokes the LLM to repair both the lemma and its proof. The details of this proof agent are provided in \Cref{sec:agent-detail}.



\begin{example}
\label{ex:formalize}
Continuing with \Cref{ex:nl-proof}, for the highlighted step in the LLM response, \approach{} generalizes and formalizes the step as lemma \texttt{power\_1}.
\begin{lstlisting}[language=Coq, basicstyle=\small\ttfamily, xleftmargin=0em,xrightmargin=0em]
(* Other lemmas omitted *)
Lemma power_1: forall x: A, power x 1 = x.
Proof.  trivial.  Qed.
\end{lstlisting}
This lemma will be further added to the context of CoqHammer during online theorem proving.
\end{example}

\subsection{Stage 2: Online Theorem Proving}
\label{sec:proving}
After extracting lemmas from the training set, given a target theorem to be proved, we insert the extracted lemmas into the context of the target theorem, thereby making them accessible to CoqHammer during the proof search. Because these additional lemmas capture the internal proof strategies of the LLM, the proving capabilities of CoqHammer are improved through the integration of LLM-extracted knowledge.


\begin{example}
\label{ex:provide}    
Continuing with \Cref{ex:formalize}, after inserting the LLM-generated lemmas into the context, we observe that CoqHammer is then able to automatically prove our target theorem~(\Cref{ex:running}). This improvement arises because, after adding the lemma \texttt{power\_1}, CoqHammer can rewrite the theorem statement of \Cref{ex:running} as
\texttt{dot (power x 1) (power x n) = dot (power x n) (power x 1)}. Using other existing theorems, this can be further reduced to \texttt{power x (1 + n) = power x (n + 1)}, which follows directly from the commutativity of addition.
\qed
\end{example}


\subsection{Implementation Details}
\label{sec:agent-detail}

We implement \approach{} on Rocq 8.20.0. To retrieve relevant type and function definitions, which are filled into the prompt described in \Cref{fig:prompt-design-nl-proof-gen}, we employ meta-programming using the OCaml libraries of Coq and develop a dedicated tactic that pretty-prints these definitions.

A natural concern is whether adding a large number of new lemmas burdens the hammer. However, CoqHammer uses a built-in KNN-based lemma selection mechanism for retrieving relevant lemmas from the context. Crucially, regardless of the total number of lemmas available, CoqHammer always retrieves a fixed number of candidate lemmas for each proof task. This design ensures that, even when many new lemmas are added to the context, the efficiency of CoqHammer remains largely unaffected.

Finally, we describe the design of our proof agent. The agent takes as input a lemma, its corresponding proof, and a limit \texttt{round\_limit}. It iteratively attempts to repair errors for at most \texttt{round\_limit} rounds, and returns the corrected versions if successful; otherwise, it reports failure once the round limit is reached. The repair process follows a module-based approach~\cite{DBLP:conf/kbse/LuD024}. We define a set of predefined error types, each associated with a dedicated LLM-assisted repair module, where each module is equipped with a specialized prompt tailored to its error category. When an error message matches one of these predefined types, the corresponding module is invoked. If the error does not match any predefined category, the default repair agent, which prompts the LLM using only the raw error message without additional guidance, is executed.

%% file: evaluation.tex
\newcommand{\approachcld}{{\approach{}{{\sc Cld}}}}
\newcommand{\approachgpt}{{\approach{}{{\sc Oai}}}}

\section{Evaluation}

\paragraph{Research Questions}
We focus on the following research questions.
\begin{itemize}
\item \textbf{RQ1:} How effectively can \approach{} extract proof strategies from LLMs?
\item \textbf{RQ2:} To what extent do the proof strategies extracted by \approach{} improve the success rate of symbolic provers?
\item \textbf{RQ3:} Does \approach{} work with other backend LLMs?
\item \textbf{RQ4:} How does the extraction method in \approach{} contribute to its overall performance?
\item \textbf{RQ5:} Do the extracted lemmas benefit LLM agents?
\end{itemize}

\paragraph{Dataset}
The evaluation of our approach requires a training set and a testing set. We follow the standard machine learning practice, collecting a dataset and using n-fold cross-validataion to evaluate our approach. We constructed a dataset of proved theorems collected from open-source Rocq projects focused on software verification. These projects include:
\begin{itemize}
\item \compcert{}~\cite{DBLP:conf/itp/KrebbersLW14}: a certified C compiler implemented in Rocq,
\item \extlib{}~\cite{coqextlib}: a community-driven Rocq library,
\item \coqart{}~\cite{bertot2013interactive}: a textbook on Rocq for building certified programming languages and data structures,
\item \vfa{}~\cite{Appel:SF3}: the third volume of Software Foundations focusing on verified functional algorithms.
\end{itemize}
For \compcert{}, due to its large number of theorems, we follow the procedure of previous works~\cite{11029818,DBLP:journals/corr/abs-2410-19940} and randomly sample 1,000 theorems. For the other projects, we collected all theorems available in each project. In total, our dataset comprises \theoremsum{} theorems from these projects. 


\paragraph{Configurations}
For generating natural language proofs and summarizing lemmas, we use Claude 3.7 Sonnet in non-thinking mode~\cite{claudesonnet}. For proof agent demonstrated in \Cref{sec:agent-detail}, we set \texttt{round\_limit} to 8.


All experiments were conducted on a server equipped with an 80-core Intel(R) Xeon(R) Gold 6230 CPU, 512 GB RAM, and running Ubuntu 20.04.6.

\subsection{\textbf{RQ1:} How effectively can \approach{} extract proof strategies from LLMs?}
\label{sec:rq1}

\paragraph{Procedure}
This experiment evaluates the offline lemma mining stage (Stage 1, \Cref{sec:gen-nl-proof,sec:formalize}) of \approach{}. We apply \approach{} to mine proof strategy lemmas with corresponding proofs across all collected 2,394 theorems. The results are summarized in \Cref{tab:ex_llm_claude-3-7}, where for each project, we further report: i) the total number of natural-language proof steps generated by the backend LLM, and ii) the number of lemmas successfully proved by our proof agent. We also report the average number of added lemmas, relative to the number of existing theorems, which reflects the average number of internal proof strategies that can be extracted by \approach{} from each theorem.

\begin{table}[ht]
\centering
\caption{Detailed Statistics for Proof Strategies Extraction}
\label{tab:ex_llm_claude-3-7}
\begin{scriptsize}
    \begin{tabular}{ccc@{\hspace{2pt}}clccl}
\toprule
& \multicolumn{4}{r}{\textit{Claude 3.7 Sonnet}} & \multicolumn{3}{c}{\textit{o3-mini}} \\
\cmidrule(lr){3-5} \cmidrule(lr){6-8}
 & \textit{\#Theorems} & \textit{\#NL Steps}  & \multicolumn{2}{c}{\textit{\#Lemmas}} & \textit{\#NL Steps}  & \multicolumn{2}{c}{\textit{\#Lemmas}} \\ 
\midrule
\compcert{} & 1000 & 3650 &  1611 & ($\times$ 1.61) & 4583 & 721 & ($\times$ 0.72)\\
\extlib{} & 187 & 468  & 300 & ($\times$ 1.60) & 439 & 243 & ($\times$ 1.30)\\
\coqart{} & 947 & 3573  & 2409 & ($\times$ 2.54)  & 3602 & 1906 & ($\times$ 2.01)\\
\vfa{} & 260  & 1023  &  263 & ($\times$ 1.01) & 1004 & 215 & ($\times$ 0.83) \\ 
\midrule
Total & 2394 &  8714 & 4583 & ($\times$ 1.91) & 9947 & 3085  & ($\times$ 1.29)\\ 
\bottomrule
\end{tabular}
\end{scriptsize}
\end{table}

\paragraph{Results}
\Cref{tab:ex_llm_claude-3-7} shows that, on average, \approach{} is able to extract 1.91 new lemmas for each original theorem. These results demonstrate that \approach{} has been effective in extracting internal proof strategies from LLM. 

\paragraph{Case Studies} We observe that about half of the NL steps are not successfully generalized and formalized as lemmas. We analyzed the main reasons, and summarize three main reasons qualitatively below. 

\smallskip\textit{Incorrect NL Proof due to Hallucinations in LLMs.} Consider the following theorem from our dataset.
\begin{lstlisting}[language=Coq,basicstyle=\small\ttfamily,xleftmargin=0em,xrightmargin=0em]
Hypothesis bad_inference := forall P Q : Prop, (P -> Q) -> ~ P -> ~ Q.
Theorem incons: False.
\end{lstlisting}
It assumes an incorrect logical rule: it swaps the premise and conclusion in the contrapositive rule, whereas the correct form should be $\mathtt{(P \rightarrow Q) \rightarrow (\sim Q \rightarrow \sim P)}$. Our goal is to show that, by applying this incorrect rule, one can derive \texttt{False} without any assumptions, thereby establishing the logical inconsistency of the incorrect inference rule.

However, when generating the natural-language proof for the theorem above, the LLM produces the following response:
\begin{lstlisting}[basicstyle=\scriptsize\ttfamily,frame=single,xleftmargin=3em,xrightmargin=3em,breaklines=true,breakindent=0pt]
...
This is actually a valid principle in intuitionistic logic! It's a contrapositive form that says: if P implies Q, and P is false, then Q must be false. Since the hypothesis is valid, we cannot prove False from it directly.
\end{lstlisting}
In this case, the LLM falls into the hallucination~\cite{DBLP:journals/tois/HuangYMZFWCPFQL25}, it misidentifies the incorrect hypothesis as the correct contrapositive rule and, as a result, refuses to provide any natural-language proof. 

\smallskip\textit{Failure in Converting an NL step into a Lemma.}
Consider the following theorem from our dataset, which concerns another fundamental property of the generalized \texttt{power} function on \texttt{Monoid} (see \Cref{fig:running-ex} for the definition).
\begin{lstlisting}[language=Coq,basicstyle=\small\ttfamily,xleftmargin=0em,xrightmargin=3em]
Lemma power_of_power_monoid: forall {M: monoid dot one} (x: A) (n p: nat),  
    power (power x n) p = power x (p * n).
\end{lstlisting}
The LLM-generated natural language proof includes the following step (highlighted in red), which involves manipulating natural numbers:
\begin{lstlisting}[basicstyle=\scriptsize\ttfamily,frame=single,xleftmargin=3em,xrightmargin=3em,breaklines=true,breakindent=0pt,escapechar=@]
...
we have power x @\texttt{\textcolor{red}{(n + (q * n))}}@, which equals power x @\texttt{\textcolor{red}{((1 + q) * n)}}@.
...
\end{lstlisting}
Ideally, we would expect the LLM to derive the following lemma, which is not present in the standard library of natural numbers.
\begin{lstlisting}[language=Coq,basicstyle=\small\ttfamily,xleftmargin=0em,xrightmargin=0em]
Lemma nat_succ_mul_custom: forall (n q: nat), n + (q * n) = (1 + q) * n.
\end{lstlisting}
However, the LLM fails to identify this lemma in the complex context involving the \texttt{power} function and natural number manipulation, and does not generate it when prompted to summarize its own natural-language proof.

\smallskip\emph{Failures in Proving a Lemma.}
We note that the effectiveness of \approach{} is also significantly influenced by the proving capabilities of the underlying proof agent. A stronger proof agent can verify a larger fraction of LLM-summarized lemmas, resulting in a more comprehensive set of extracted proof strategies. The development of advanced proof agents is orthogonal to this paper. In the future, equipping \approach{} with a more powerful agent could yield an even richer library of internal proof strategies derived from LLMs.

\begin{rqbox}
\textbf{Answer to RQ1:} 
\approach{} is overall effective in extracting internal proof strategies from the LLM via a training set. 

\end{rqbox}



\begin{table*}[ht]
\centering
\caption{Performance of CoqHammer with and without Lemmas Generated By \approach{}}
\label{tab:ex_sym_claude3.7}
\begin{scriptsize}
\begin{tabular}{ccr@{\hspace{2pt}}lcr@{\hspace{2pt}}l}
\toprule
 & \multicolumn{3}{c}{\textit{Claude 3.7 Sonnet}} & \multicolumn{3}{c}{\textit{o3-mini}} \\
 \cmidrule(lr){2-4} \cmidrule(lr){5-7}
 & CoqHammer & \multicolumn{2}{c}{\begin{tabular}[c]{@{}c@{}}CoqHammer\\+\approach{}\end{tabular}} & CoqHammer & \multicolumn{2}{c}{\begin{tabular}[c]{@{}c@{}}CoqHammer\\+\approach{}\end{tabular}} \\ 
 \midrule
\compcert{} & 293 &  \textbf{345} & ($\uparrow$ \textbf{17.75\%}) & 293 & \textbf{332} & ($\uparrow$ \textbf{13.31\%})   \\ 
\extlib{} & 78 & \textbf{82} & ($\uparrow$ \textbf{5.13\%}) & 78 & \textbf{82} & ($\uparrow$ \textbf{5.13\%}) \\ 
\coqart{} & 446 & \textbf{501} & ($\uparrow$ \textbf{12.33\%}) & 446 & \textbf{504} & ($\uparrow$ \textbf{13.00\%})   \\ 
\vfa{} & 63 &  \textbf{70} & ($\uparrow$ \textbf{11.11\%}) & 63 &  \textbf{69} & ($\uparrow$ \textbf{9.52\%})   \\ 
\midrule
Total & 880 & \textbf{998} & ($\uparrow$ \textbf{13.41\%}) & 880 & \textbf{987} & ($\uparrow$ \textbf{12.16\%})   \\ 
\bottomrule
\end{tabular}
\end{scriptsize}
\end{table*}

\subsection{ \textbf{RQ2: } Does the proof strategies extracted by \approach{} enhance the capabilities of symbolic provers?}



\paragraph{Procedure}
This experiment evaluates the online theorem proving stage (Stage 2, \Cref{sec:proving}) of \approach{}.
We compare the performance of CoqHammer with and without access to the lemmas mined in the previous research question~(\Cref{sec:rq1}).
We apply the standard 3-fold cross-validation for this evaluation. Specifically, we partition the collected theorems into three disjoint parts with equal size, and conduct three separate experiments. In the $i$-th experiment ($1 \leq i \leq 3$), the $i$-th part serves as the set of theorems to be proved, while the remaining two parts serve as the training set. 

For each experiment, we use the default configuration of CoqHammer. Specifically, the timeout is set to 20 seconds, and the number of lemmas retrieved from the context is fixed at 1024.

\paragraph{Results}
The comparison results are presented in Table~\ref{tab:ex_sym_claude3.7}. 
In terms of the number of proved theorems, \approach{} achieves an improvement ranging from 5.13\% to 17.75\% over CoqHammer across different projects. Overall, it proves \evalres{} more theorems than CoqHammer.


Notably, \approach{} achieves its greatest improvement on \compcert{}, a real-world certified compiler, demonstrating the effectiveness of \approach{} in practical Rocq proof engineering.

\paragraph{Case Studies} Below, we present some case studies to have a more fine-grained understanding of how \approach{} enhances CoqHammer.

\emph{Avoiding Induction.} Automated induction remains a long-standing challenge for symbolic provers~\cite{fm24,calm}. Notably, CoqHammer does not support induction at all~\cite{DBLP:journals/jar/CzajkaK18}. However, incorporating additional lemmas can help symbolic provers circumvent the need for induction. Below, we present a concrete example.

Continuing with our running example (\Cref{ex:formalize}; see also \Cref{fig:running-ex} for details), the target theorem \texttt{power\_commute\_with\_x} requires induction over \texttt{n} and, as a result, cannot be proved by CoqHammer. However, by adding the lemma \texttt{power\_1}, the proof can be reduced to the commutativity of natural numbers--an established result in the Rocq standard library. This avoids the need for induction and enables CoqHammer to successfully prove the target theorem.

Moreover, upon further inspection of the experimental results, we find that 42.52\% of the improvement achieved by \approach{} involves theorems that were originally proved by the developer using induction.

\emph{More Robust Lemma Selection.} The problem of identifying the set of useful lemmas is another fundamental challenge for symbolic provers~\cite{prorokovic2021improving, jiang2022thor, mikula2024magnushammer}. Under this context, we also find that \approach{} enhances the robustness of lemma selection procedure of CoqHammer. Below we present an example. 
\begin{lstlisting}[language=Coq,basicstyle=\small\ttfamily]
Theorem target: forall x (i: interval), In x i -> ~empty i.
\end{lstlisting}
Here, \texttt{i} is an interval of integers and \texttt{x} is an integer. The theorem asserts that if there exists a number in the interval, then the interval is non-empty. Although the context consists of the contrapositive of the target theorem, see the lemma \texttt{contra} below, CoqHammer does not select this lemma and fails to prove the target theorem.
\begin{lstlisting}[language=Coq,basicstyle=\small\ttfamily]
Lemma contra: forall x (i: interval), ~empty i -> In x i.
\end{lstlisting}
Fortunately, after training \approach{} on \texttt{contra}, it generates an equivalent form of the lemma \texttt{contra}, as shown in \texttt{contra\_2} below. This lemma is successfully selected by CoqHammer, enabling it to prove the target theorem.
\begin{lstlisting}[language=Coq,basicstyle=\small\ttfamily,xleftmargin=0em]
Lemma contra_2: forall x (i: interval), empty i -> In x i -> False.
\end{lstlisting}

\begin{rqbox}
\textbf{Answer to RQ2:} 

\approach{} significantly enhances the performance of CoqHammer in terms of the number of theorems proved.
\end{rqbox}

\subsection{\textbf{RQ3:} Does \approach{} work with other backend LLMs?}

\paragraph{Procedure} This experiment aims to understand whether \approach{} is specific to a particular backend LLM, or can generically extract strategies from different LLMs. We replace Claude 3.7 Sonnet--the backend LLM originally used in RQ1 and RQ2--with OpenAI o3-mini~\cite{opanaio3mini}, and rerun the experiments described in RQ1 and RQ2.

\paragraph{Results} The results are summarized in \Cref{tab:ex_llm_claude-3-7} and \Cref{tab:ex_sym_claude3.7}. 
We observe that \approach{} remains effective in producing formalized lemmas in Rocq that reflect the internal proof strategies of o3-mini. 
Furthermore, replacing the backend LLM with o3-mini continues to enhance the proving capability of CoqHammer, increasing the number of proved theorems by 12.16\% and the number of tactics automated by 26.27\%.

Since the two sets of experiments extract internal strategies from different LLMs, we are also interested in whether the lemmas extracted from two LLMs together could further improve the performance. We make a union of the two sets of lemmas, and evaluate how this set improves CoqHammer. The result shows that CoqHammer proves 1019 theorems, which is higher than 998 theorems (from Claude 3.7 Sonnet) and 987 theorems (from o3-mini). This suggests that the internal strategies from different LLMs are not fully overlapped and may complement each other.

\begin{rqbox}
\textbf{Answer to RQ3:}
\approach{} is not specific to a particular backend LLM and can extract internal strategies from different backend LLMs.
\end{rqbox}

\subsection{RQ4: How does the NL proof generation in \Cref{sec:gen-nl-proof} Contribute to the Effectiveness of \approach{}?}


\paragraph{Procedure}  This experiment evaluates the impact of the NL proof generation step (\Cref{sec:gen-nl-proof}). Instead of using the NL proof generation step, we directly ask the LLM to generate formalized lemmas.


\paragraph{Results} As shown in \Cref{tab:ablations}, the result indicate a 63.56\% decrease in the number of extracted lemmas and a 6.21\% decrease in performance.

\begin{table}[H]
	\centering
	\caption{An ablation study of the extraction method}
	\label{tab:ablations}
\begin{scriptsize}
\begin{tabular}{ccccc}
\toprule
 & \multicolumn{2}{c}{\textit{Original}} & \multicolumn{2}{c}{\textit{-NL}} \\
 \cmidrule(lr){2-3} \cmidrule(lr){4-5}
 & \textit{\#Lemmas} & \textit{\#Proved Theorems} & \textit{\#Lemmas}   &  \textit{\#Proved Theorems} \\ 
 \midrule
\compcert{} & 1611  & 345 & 212 ($\downarrow \textbf{86.84\%}$)  &308 ($\downarrow \textbf{10.72\%}$)  \\ 
\extlib{} & 300 & 82 & 137 ($\downarrow \textbf{54.33\%}$) & 78 ($\downarrow \textbf{4.88\%}$)  \\ 
\coqart{} & 2409 & 501 &1263 ($\downarrow \textbf{47.57\%}$) & 484 ($\downarrow \textbf{3.39\%}$)   \\ 
\vfa{} & 263 & 70  & 58 ($\downarrow \textbf{77.95\%}$) & 66 ($\downarrow \textbf{5.71\%}$)   \\ 
\midrule
Total & 4583  & 998 &  1670 ($\downarrow$ \textbf{63.56\%}) & 936 ($\downarrow \textbf{6.21\%}$)    \\ 
\bottomrule
\end{tabular}
\end{scriptsize}
\end{table}

\begin{rqbox}
\textbf{Answer to RQ4:}
The NL proof generation step in \approach{} is essential for effectively extracting strategies from LLMs and, in turn, enhancing the proving capabilities of CoqHammer.
\end{rqbox}

\subsection{RQ5: Can \approach{} enhance the capability of LLM agents?}
Currently, many approaches~\cite{DBLP:conf/kbse/LuD024,11029818,DBLP:conf/iclr/WangXZLCHXSX0LL24} employ LLM agents for automated theorem proving, where an agent interacts with LLMs, symbolic provers, and proof assistants to construct proofs. While our primary objective is to enhance purely symbolic provers, it is natural to ask whether the lemmas extracted by \approach{} can also improve the performance of LLM agents. 

\paragraph{Procedure} To investigate this, we implement an LLM agent on top of the agent described in Section~\ref{sec:agent-detail} by incorporating a retrieval-augmented generation (RAG) module~\cite{DBLP:conf/nips/LewisPPPKGKLYR020}. The RAG module retrieves relevant lemmas from all lemmas accessible in the current context to assist the LLM in generating proofs. When the extracted lemmas are inserted into the context, they become available for retrieval by the RAG module. Furthermore, our agent can also be configured to enable or disable CoqHammer in proof, and will always try CoqHammer first when it is enabled. We evaluate two settings of the agent:
\begin{itemize}
\item \textbf{Setting I:} The agent has access to the LLM, CoqHammer, and the proof assistant. We compare its performance with and without the lemmas mined by \approach{} inserted into the context.
\item \textbf{Setting II:} The agent has access to the LLM and the proof assistant only (CoqHammer disabled). Again, we compare its performance with and without the lemmas mined by \approach{}.
\end{itemize}

Setting I directly addresses the research question posed in this section, while Setting II serves as an ablation study that provides a more fine-grained understanding of the contribution of the extracted lemmas. In both settings, we evaluate the number of theorems successfully proved and the total number of tokens consumed. The LLM choice retains the same as \Cref{sec:rq1}.

\begin{table}
	\centering
	\caption{Performance of the LLM Agents}
	\label{tab:llm_agent}
\begin{scriptsize}
		\begin{tabular}{ccccc}
		\toprule
		\multicolumn{1}{l}{}  & \multicolumn{1}{l}{\begin{tabular}[l]{@{}c@{}}LLM Agent\\~+ CoqHammer\end{tabular}}  & \multicolumn{1}{r}{\begin{tabular}[c]{@{}c@{}}LLM Agent \\+ CoqHammer \\+ \approach{} \end{tabular}}  & LLM Agent  & \multicolumn{1}{l}{\begin{tabular}[c]{@{}c@{}}LLM Agent\\~+ \approach{}\end{tabular}}  \\
		\midrule
		\compcert{}   &    502    & 545  ($\uparrow \textbf{8.57\%}$)   & 363 & 406  ($\uparrow \textbf{11.85\%}$)     \\
		\extlib{}       &   104         & 106 ($\uparrow \textbf{1.92\%}$)   & 76 & 77     ($\uparrow \textbf{1.32\%}$)  \\
		\coqart{}    &  667         &  688   ($\uparrow \textbf{3.15\%}$)  & 584  & 581       ($\downarrow \textbf{0.51\%}$)    \\
		\vfa{}    &     128        & 118  ($\downarrow \textbf{7.81\%}$)  & 105  & 94  ($\downarrow \textbf{10.47\%}$)      \\
        \midrule
        Total   &  1401 & 1457 ($\uparrow \textbf{4.00\%}$)  & 1128 & 1158 ($\uparrow \textbf{2.65\%}$)  \\
        \midrule
        Tokens  & 1.379$\times 10^7$ & 1.298$\times 10^7$ ($\downarrow \textbf{5.87\%}$) & 1.798$\times 10^7$ & 1.727$\times 10^7$ ($\downarrow \textbf{3.94\%}$) \\
        \bottomrule                                                                              
	\end{tabular}
\end{scriptsize}
\end{table}

\paragraph{Results} The results are presented in \Cref{tab:llm_agent}. Comparing the column 2 and column 3, we observe an improvement of 4.00\% in Setting I. To understand why this improvement happens, we can compare the latter two columns, which show an improvement of 2.65\% with LLM proof generation alone in Setting II. Although these lemmas represent knowledge internal to the LLM, the RAG component remains important for surfacing and prioritizing this knowledge during proof generation of LLM. Notably, the improvement in Setting I exceeds that of Setting II, indicating that deeper integration between CoqHammer and the LLM strategies proves lemmas that the agent cannot. 
This is because, as we conjecture, LLM only invokes CoqHammer as a blackbox in a standard agent, but our approach allows to integrate the strategies into the internal proof search process in CoqHammer. 

We also observe reduced token usage in both settings. This reduction arises because the enhanced CoqHammer can solve more theorems independently, thereby decreasing the number of LLM calls required during proof generation.

\begin{rqbox}
\textbf{Answer to RQ5:}
\approach{} improves the performance of LLM agents, yielding both stronger proving capability and reduced token usage. 
\end{rqbox}

%% file: related.tex
\section{Related Work}
\label{sec:related}

\paragraph{Machine Learning for Theorem Proving}
There is a long line of research focused on applying machine learning techniques to theorem proving~\cite{copra,DBLP:conf/icml/YangD19,DBLP:journals/pacmpl/FirstBG20,DBLP:conf/icse/FirstB22,DBLP:conf/pldi/Sanchez-SternAS20,DBLP:conf/icml/BlaauwbroekORMP24,DBLP:conf/mkm/BlaauwbroekUG20,DBLP:conf/sigsoft/FirstRRB23,DBLP:conf/iclr/WangXZLCHXSX0LL24,DBLP:journals/corr/abs-2410-19940,DBLP:conf/kbse/LuD024,11029818,prorokovic2021improving,jiang2022thor,mikula2024magnushammer}. These approaches can be divided into two categories: i) Training a model for theorem proving. ii) Employing an agentic approach that iteratively generates proofs through interaction with LLMs.

All of these methods require online interaction with trained models during the process of proving new theorems, which can be both expensive and insecure, as discussed in \Cref{sec:introduction}. In contrast, \approach{} only invokes the symbolic prover when proving new theorems, making the process secure, cost-effective, and suitable for execution on a personal computer.

\paragraph{Knowledge Extraction from LLMs}
This line of research focuses on extracting the internal knowledge of LLMs into alternative representations. Previous work in this area has primarily concentrated on extracting LLM knowledge into knowledge graphs~\cite{DBLP:conf/emnlp/ZhangS24,DBLP:journals/corr/abs-2403-08345}.

However, these approaches are not directly applicable to our setting, as knowledge graphs 
capture a set of concepts and their relations, does not represent the internal proof strategies of LLMs, and is not accessible to symbolic provers.
In contrast, \approach{} extracts the internal proof strategies of LLMs as verified lemmas in Rocq, making them accessible to CoqHammer.

\paragraph{Symbolic Provers}
There have been numerous efforts to develop symbolic provers for various proof assistants, such as CoqHammer~\cite{DBLP:journals/jar/CzajkaK18} for Rocq~\cite{CoqRefMan8.20}, Lean-auto~\cite{qian2025leanautointerfacelean4} for Lean4~\cite{DBLP:conf/cade/MouraKADR15}, and Sledgehammer~\cite{DBLP:conf/cade/BohmeN10} for Isabelle~\cite{DBLP:books/sp/NipkowPW02}. Currently, \approach{} implements strategy extraction on top of Rocq and has demonstrated significant improvements over CoqHammer. Our approach can be adapted to other symbolic provers by translating the formalized theorems from Rocq to other proof assistants, which we leave as future work.

%% file: discussion.tex

\section{Conclusion and Future Work}
In this paper, we explore a new research direction: leveraging the internal knowledge of LLMs to augment symbolic provers. 
We introduce \approach{}, a framework that extracts proof strategies from LLMs and formalizes them as lemmas in Rocq. 
Empirical results demonstrate that, on open-source Rocq projects for software verification, \approach{} increases the success rate of CoqHammer by \evalres{}.


We would also like to point out that our work is also valuable in understanding how LLMs work, a fundamental research question regarding LLMs. 
A standard approach towards this question is to build various interpretable symbolic models to approximate the behavior of the LLMs.
Our approach provides a means to approximate an LLM in a symbolic prover. If, with potential future work, the enhanced symbolic prover could achieve a similar performance with LLMs, this provides a potential explanation of how LLMs conduct formal proofs.



